\documentclass[12pt,preprint]{aastex}   %{/home/rms/text/aastex}
\usepackage{psfig,natbib}

\def\feka{Fe K$\alpha$}
\def\chandra{{\it Chandra}} 
\def\xmm{{\it XMM-Newton}} 
\def\asca{{\it ASCA}} 
\def\hst{{\it HST}} 
 
\def\sax{{\it BeppoSAX}} 
 
\def\rosat{{\it ROSAT}}

\def\lum{erg s$^{-1}$}
\def\flux{erg cm$^{-2}$ s$^{-1}$}
\def\nh{cm$^{-2}$}
\def\arcsec{$^{\prime\prime}$}
\def\arcmin{$^{\prime}$}
\def\deg{$^{\circ}$}

\def\ltsima{$\; \buildrel < \over \sim \;$}
\def\simlt{\lower.5ex\hbox{\ltsima}} % < over ~
\def\gtsima{$\; \buildrel > \over \sim \;$}
\def\simgt{\lower.5ex\hbox{\gtsima}} % > over ~

\begin{document}

\title{The XMM-Newton View of the Nucleus of NGC~4261}

\author{R.M. Sambruna\altaffilmark{\;1}, 
M. Gliozzi\altaffilmark{\;1}, M. Eracleous\altaffilmark{\;2}, 
W.N.Brandt\altaffilmark{\;2}, and R. Mushotzky\altaffilmark{\;3}}

\altaffiltext{1}{George Mason University, Dept. Of Physics \&
Astronomy, 4400 University Drive, Ms 3f3, Fairfax, Va 22030 (e-mail:
rms,mario@physics.gmu.edu).}

\altaffiltext{2}{The Pennsylvania State University, Department of
Astronomy And Astrophysics, 525 Davey Lab, State College, Pa 16802
(e-mail: mce,niel@astro.psu.edu).}

\altaffiltext{3}{NASA Goddard Space Flight Center, Code 662, 
Greenbelt, MD 20771 (e-mail: richard@xray-5.gsfc.nasa.gov).}

\clearpage
\begin{abstract}

We present the first results from an \xmm\ observation of the FRI
galaxy NGC~4261, which harbors a supermassive black hole and a
low-ionization nuclear emission-line region (LINER). Here we focus on
the X-ray properties of the nucleus, using the EPIC pn data. The
0.6--10 keV continuum in best fitted by a thermal component with $kT
\sim 0.7$ keV, plus a power law with photon index $\Gamma \sim 1.4$,
absorbed by a column density N$_H \sim 4 \times 10^{22}$ \nh. An
unresolved Fe K emission line with EW $\sim$ 280 eV is detected at
$\sim$ 7 keV. We also detect, for the first time, short-term flux
variability from the nucleus, on a timescale of 3--5 ks. The
short-term variations rule out an ADAF as the {\it only} production
mechanism of the X-ray continuum. Instead, we argue that the inner jet
contributes to the emission in the X-ray band.

Subject Headings: galaxies: active; galaxies: jets; galaxies:
elliptical and lenticular, cD; X-rays: galaxies; galaxies: individual (NGC~4261, 
3C~270) 

\end{abstract} 

\section{Introduction}

In recent years, it was established that many nearby galaxies contain
supermassive black holes (e.g., Gebhardt et al. 2000). However, weak
or no activity is usually detected from their cores, in the optical or
X-ray band (Loewenstein et al. 2001; Kormendy \& Ho 2000). This is
puzzling, as at least in early-type galaxies large reservoirs of gas
to feed the central black holes are available in the form of hot,
diffuse galaxy halos (e.g., Fabian \& Canizares 1988).

The nearby ($\sim$ 41 Mpc) giant elliptical NGC~4261/3C270 contains a
supermassive black hole of known mass, $M_{\rm BH}\sim 5\times10^8
M_{\odot}$ (Ferrarese, Ford, \& Jaffe 1996). It is one of a handful of
early-type galaxies where low-luminosity nuclear activity is
detected. In the radio band, it exhibits a Fanaroff-Riley type I (FRI;
Fanaroff \& Riley 1974) morphology, with a compact core and twin jets
extending E-W, oriented at $i=(63 \pm 3)$\deg\ with respect to the
line of sight (Piner, Jones, \& Wehrle 2001). Optical spectroscopy
reveals a Low-Ionization Nuclear Emission-Line Region (LINER; Heckman
1980) in the nucleus of the galaxy (Ho, Filippenko, \& Sargent
1995). In the X-ray band, diffuse thermal emission with $kT
\sim 0.6$ keV is present at low energies in \rosat\ images (Worrall \&
Birkinshaw 1994). At hard X-rays, a point source is detected at the
position of the radio core in a recent \chandra\ image; its continuum
is described by a power law with photon index $\Gamma_{2-10~keV} \sim
1.3$, excess column density N$_H \sim 6 \times 10^{22}$ \nh, and
intrinsic luminosity L$_{2-10~keV} \sim 10^{41}$ \lum\ (Chiaberge et
al. 2002).  Interestingly, an unresolved narrow ($\sigma=0.05$ keV) Fe
K line was marginally detected at 6--7 keV with \asca\ (Terashima et
al. 2002; Sambruna, Eracleous, \& Mushotzky 1999).

NGC~4261 exhibits significantly different optical and X-ray properties
from the more luminous, broad-lined radio-loud sources (e.g., Sambruna
et al.  1999 and discussion therein). It was indeed suggested that the
accretion process in NGC~4261 proceeds through an Advection-Dominated
Accretion Flow (Ho 1999), or that beamed emission from the jet
dominates the energy production in this source (Chiaberge et al. 2002
and references therein). NGC~4261 qualifies as an important link to
understand the transition from low- to high-power radio sources.

To study its nuclear X-ray properties in more detail, we acquired an
\xmm\ observation of NGC~4261. In this Letter, we present a study of
the nuclear properties using the more sensitive EPIC pn data. The
archival \chandra\ image is also used to constrain the spatial
properties of the galaxy within the EPIC extraction region. A more
detailed study of the \xmm\ data will be presented in Gliozzi et
al. (2003).  The observations and the data reduction are described in
\S~2. In \S~3 we give the results and discuss their
significance. Throughout the paper we use a Friedman cosmology with
$H_0=75~{\rm km~s^{-1}~Mpc^{-1}~and}~q_0=0.5$.

\section{Observations And Data Reduction}

\begin{figure}
\psfig{figure=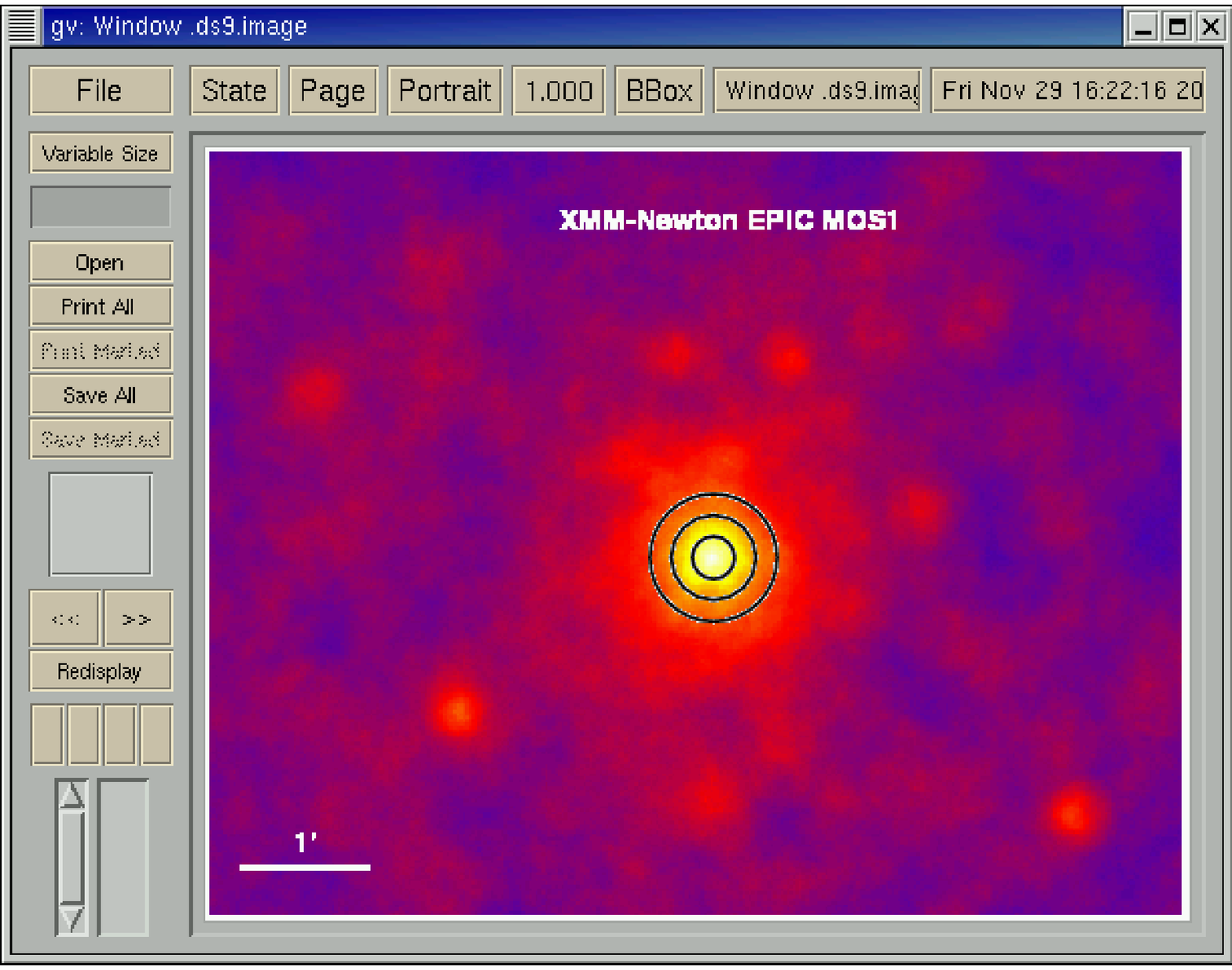,height=8.cm,width=11cm,%
bbllx=102pt,bblly=183pt,bburx=587pt,bbury=564pt,angle=0,clip=}
\psfig{figure=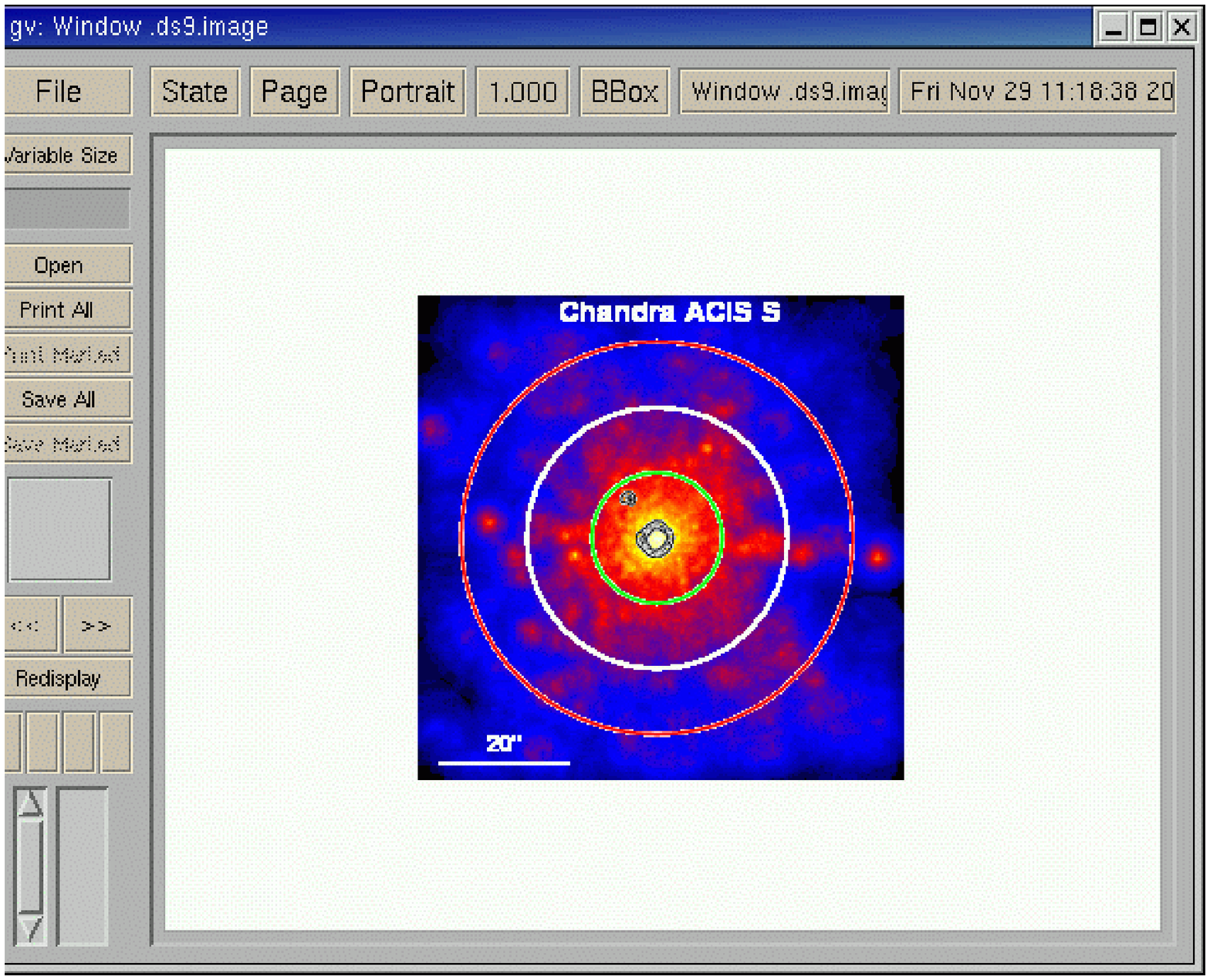,height=8cm,width=8cm,%
bbllx=210pt,bblly=252pt,bburx=455pt,bbury=496pt,angle=0,clip=}
\caption{{\it Top panel:} \xmm\ EPIC MOS1(CCD1) smoothed image of the    
NGC~4261 field in the 0.3--10 keV energy range. The black circles
indicate the extraction regions for the temporal and spectral
analyses, with radii 10\arcsec, 20\arcsec, and 30\arcsec,
respectively. {\it Bottom panel:} \chandra\ ACIS-S3 smoothed image of
NGC~4261 in the 0.2--2 keV energy range, with the hard X-ray
(2--8 keV) contours overlaid. The image was obtained applying an
adaptively smoothing procedure to a subsampled image, binned at 0.3 of
the original pixels.  The circles indicate the EPIC pn extraction
regions for the temporal and spectral analyses (10\arcsec, 20\arcsec,
and 30\arcsec). North is up and East to the left. 
\label{figure:im1}}
\end{figure}

We first used an archival 34 ks \chandra\ observation, obtained on
2000 May 6, to quantify the spatial properties of the source within
the \xmm\ extraction radius.
Figure~\ref{figure:im1} (bottom panel) shows the \chandra\
ACIS-S3 image from 0.2--2 keV, with the hard X-ray (2--10 keV)
contours overlaid.  
X-ray emission from the core, the E-W jet, the diffuse galaxy halo,
and several point sources are apparent (Gliozzi et
al. 2003, Chiaberge et al. 2002). Superposed on the
\chandra\ image are the EPIC extraction radii of 10\arcsec, 20\arcsec,
and 30\arcsec. Within 10\arcsec, only one point source
(RA=$12^h19^m23.50^s$, DEC=$05^o$49\arcmin36\arcsec) is present; its
contribution to the total X-ray flux is negligible, amounting to less
than 2\%.  At larger radii, the relative contribution of the halo,
resolved jet, and point sources increases.  A detailed
energy-dependent spatial analysis of the ACIS image will be presented
in Gliozzi et al. (2003). Here, we only note that a point source at
the position of the nucleus is present at ultra-soft energies,
0.3--0.8 keV, together with faint diffuse emission. The diffuse halo
contribution is strongest at intermediate energies, 0.8--2 keV, where
its spectrum peaks. 

The \xmm\ observation of NGC~4261 was carried out on 2001 October 16
(revolution 370). Only half of the proposed exposure was awarded, for
a total of 30~ks.  Here we report the results from the EPIC CCDs
mainly from the pn camera, because of its higher sensitivity. The
recorded events were reprocessed and screened with the latest
available release of the
\xmm\ Science Analysis Software (SAS 5.3.3), retaining only events
corresponding to patterns 0--12 (singles, doubles, triples and
quadruples). After removing periods of background flares, the net
exposure is $\sim 21$ ks. In an extraction radius of 10\arcsec, the
EPIC pn count rate in the 0.6--10 keV band is
$(1.76\pm0.01)\times10^{-1}$ counts s$^{-1}$.

The source data were extracted from circular regions of different
radii ranging from 10\arcsec\ to 30\arcsec\ (Figure~\ref{figure:im1}). 
A radius of 30\arcsec\ corresponds to the maximum value before
reaching the edge of chip 4 in the pn camera. A radius of 10\arcsec\
was finally chosen for the spectral analysis of the EPIC pn data
because it minimizes the contribution of the extended components and
point sources. A radius of 30\arcsec\ was instead used for the light
curves in order to have enough photons for the timing analysis at hard
X-rays (none of the point sources included in this radius is
demonstrably variable in the \chandra\ observation). For each
extraction radius, the count rate was corrected using the encircled
energy function and point spread function implemented in the {\tt SAS}
task {\tt arfgen}. Background data were extracted from source-free
circular regions (r=30\arcsec) on the same chip containing the central
source.  There are no signs of pile-up according to the {\tt SAS} task
{\tt epatplot}.

\section{Results and Discussion}

\subsection{Detection of Flux Variability} 

\begin{figure}
\psfig{figure=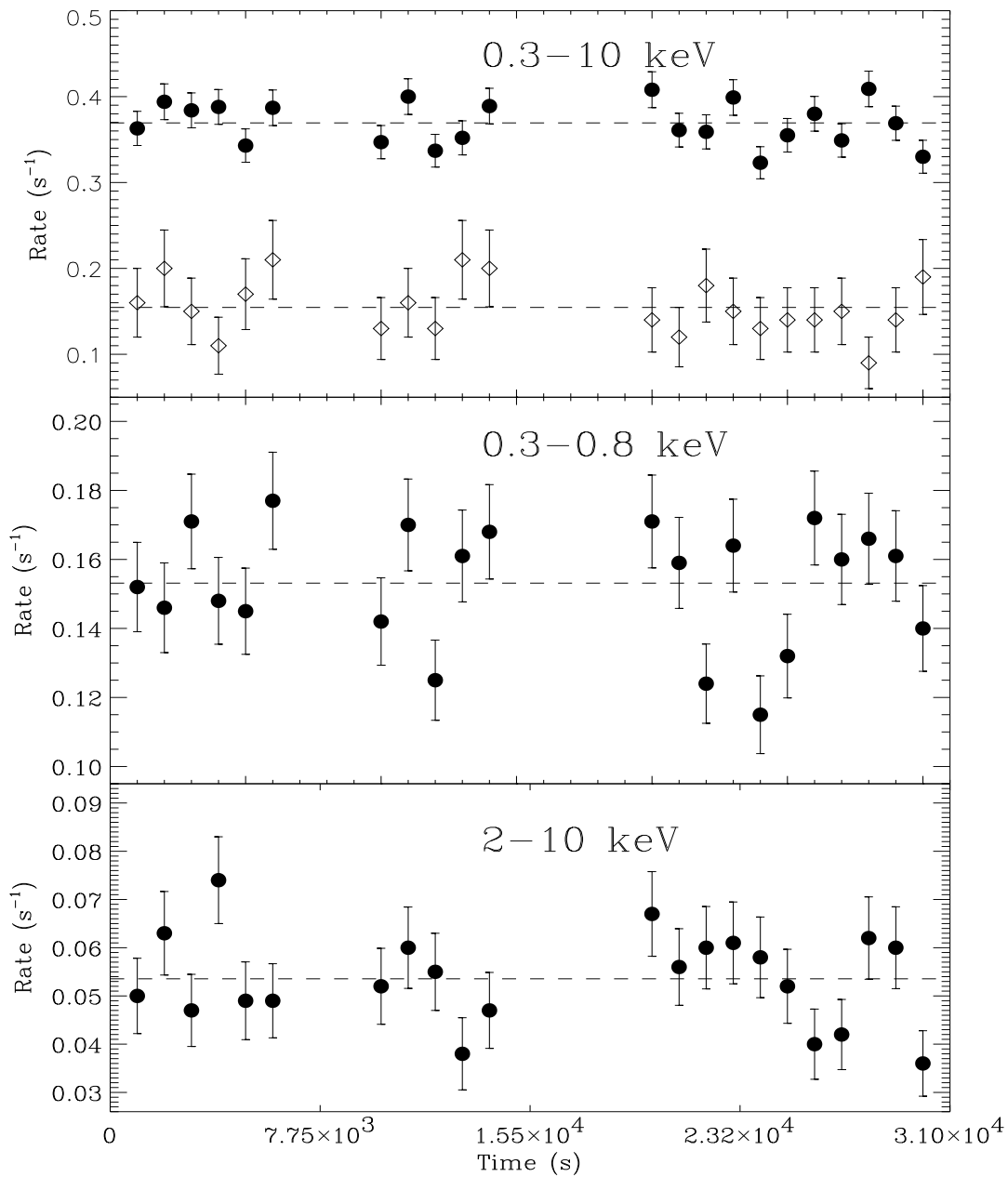,height=18cm,width=16cm,%
bbllx=65pt,bblly=20pt,bburx=520pt,bbury=570pt,angle=0,clip=}
\caption{\xmm\ EPIC pn light curves of NGC~4261 in the full 
band (top panel), ultra-soft band (middle panel), and hard band
(bottom panel). The extraction radius is 30\arcsec; time bins are 1000
s. The dashed lines indicate the average values of the source count
rate at various energies. The background light curve is also plotted
for comparison in the top panel, amplified by a factor of 10. 
\label{figure:lc1}}
\end{figure}

The 0.3--10 keV EPIC pn light curve from an extraction radius of
30\arcsec\ is shown in Figure~\ref{figure:lc1}. Low-amplitude flux
variability is present, with flux excursions of a factor 1.3
(maximum/minimum) on timescales of $\sim$ 3 ks. A $\chi^2$ test yields
a probability that the flux is constant of only P$_{\chi^2}$=2.5\%. To
verify that the variability originates in the nucleus, light curves
were extracted from 20\arcsec\ and 10\arcsec\ radii. The $\chi^2$
probability increases with increasing extraction radius of the light
curves, indicating a progressively larger dilution of the light curve
by non-nuclear light in the larger extraction radii, and confirming
that the variability is indeed of nuclear origin. We identify the
source of the X-ray variations with the nuclear point source detected
in the \chandra\ image.  While no flux variability is present in the
\chandra\ light curves of the nucleus (within 2\arcsec\ extraction
radius), the uncertainties are very large.

We investigated possible energy-dependences of the flux variations in
the EPIC data splitting the total 0.3--10 keV band into sub-bands
containing enough counts to perform the $\chi^2$ test. We find
significant flux changes at hard (2--10 keV) and ultra-soft (0.3--0.8
keV) energies, with P$_{\chi^2}$=6\% and 0.3\%, respectively. As shown
in Figure~\ref{figure:lc1}, in both energy bands the flux changes by a
factor 1.5 on timescales $\sim$ 3--5 ks (minimum to maximum) but with
different variability patterns. At intermediate energies, 0.8--2 keV,
no flux variations are detected in the EPIC light curve
(P$_{\chi^2}$=75\%), due to the presence of the spatially-extended
thermal component that dilutes the nuclear light. Note that a
non-thermal component is present in the spectrum at ultra-soft
energies (see below).

Erratic spectral variations are also present in the plot of the
hardness ratio, 0.3--0.8 keV/2--10 keV, versus time (P$_{\chi^2}
\sim$ 0.8\%). However, there are not enough counts to perform a
detailed spectral analysis as a function of time. 

The detection of short-term flux variations with EPIC poses a
challenge for an origin of the X-ray continuum in an Advection
Dominated Accretion Flow (ADAF), the favored accretion-flow structure
in low-luminsity AGNs and accretion-powered LINERs (e.g., Ho 1999). In
an ADAF, the X-ray emission comes from a large volume and relatively
long (\gtsima 1 day) variability timescales are expected, as indeed
observed with \asca\ (Terashima et al. 2002; Ptak et al. 1998).  More
specifically, the light-crossing time of an ADAF of radius $r$ in
NGC~4261 is given by $\tau = 2.5\times 10^5\;\left({r\over 100\, r_{\rm g}}\right)
\left({M\over 5\times 10^8\; {\rm M}_{\odot}}\right)$ s,
%\begin{equation}
%\tau = 2.5\times 10^5\;\left({r\over 100\, r_{\rm g}}\right)
%\left({M\over 5\times 10^8\; {\rm M}_{\odot}}\right)~s\; ,
%\end{equation}
where $r_{\rm g} \equiv GM/c^2$ is the gravitational radius of the
black hole. Since this time scale is about 2 orders of magnitude
longer than the observed variability time scale, it is unlikely that
the variable component of the X-ray band originates in an ADAF.

The variable component of the X-ray flux in NGC~4261 is more likely to
be associated with the inner X-ray jet. 
%To explore this possibility,
%we compared the light-crossing time of the base of the jet (as
%inferred from radio observations) to the observed X-ray variability
%timescale. The jet inclination of 63\deg\ and intrinsic plasma speed
%of $\beta=0.46$ on pc scales (Piner et al. 2001) imply a Doppler
%factor $\delta \sim 1.4$. The EPIC timescale of 5 ks thus implies an
%intrinsic size for the emitting region of $\sim 2.5 \times 10^{14}$
%cm, or a jet opening angle (assuming a cone) of $\sim$ 2\deg\ at
%$\sim$ 1 pc from the nucleus. This is consistent with the upper limit
%for the jet opening angle from high-resolution radio observations,
%\ltsima 5\deg\ (Jones et al. 2001), suggesting that emissivity
%variations across a quasi-cylindrical structure can lead to the
%observed X-ray flux variations. 
However, the detection of a relatively strong Fe K line (see below)
indicates that jet emission is not dominant over the entire observed
X-ray band; another continuum production mechanism must be
present. Alternatively, the X-ray continuum is produced in the jet but
is only mildly beamed. This is the case if the jet has a fast spine
and slower walls (e.g., Chiaberge et al. 2000 and references therein), 
and the observed X-ray continuum is produced by the decelerated plasma.
%In fact, jet emission is
%arguably not dominant in {\it any} portion of the observed band, based
%on the following argument: if we take the jet emission to be 
%intrinsically variable by approximately a factor of 2 (as observed in
%blazars), then the fact that we observe variations of the order of 1.3
%implies that the variable jet emission contributes only 40\% of the
%total flux.

\subsection{Excess X-ray Absorption in the Nucleus of NGC~4261} 

At first we focused the spectral analysis on the time-averaged EPIC pn
spectrum in the 0.6--10 keV range, where the instrument is best
calibrated, with an extraction radius of 10\arcsec\ (see
above). Following indications from previous X-ray observations, we
first fitted the spectrum with a thermal model (\verb+apec+ in
\verb+XSPEC+ v.11) plus a power law. The column density, obscuring
both components, was fixed at the Galactic value of N${_H}=1.52\times
10^{20}$ \nh\ (Dickey \& Lockman 1990). The resulting fit is formally
acceptable ($\chi^2=169$ for 136 degrees of freedom). However, the
resulting photon index ($\Gamma\sim 0.32$) is flatter than in other
LINERs and weak-line radio galaxies like NGC~4261 (e.g., Terashima et
al. 2002; Sambruna et al. 1999).

\begin{figure}
\psfig{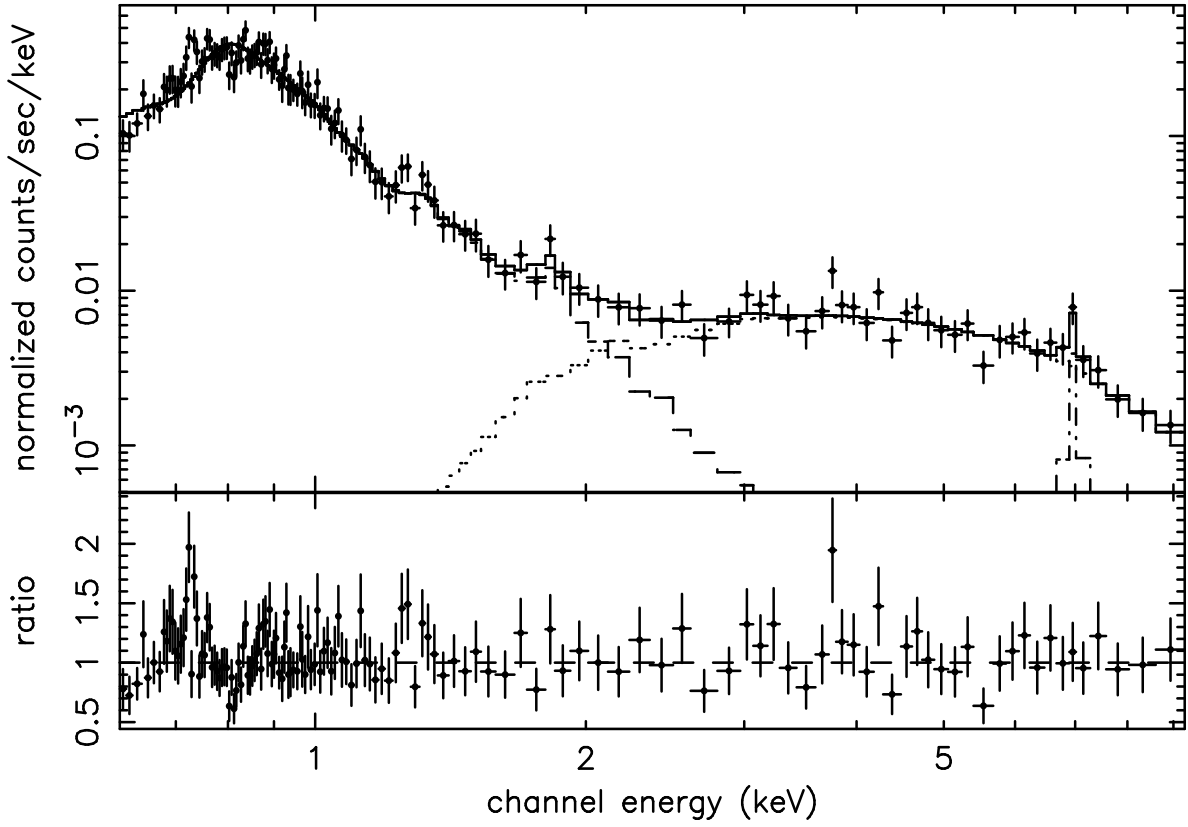}
\caption{XMM-Newton EPIC pn spectrum of the central source fitted with
a Raymond-Smith plus an absorbed power-law plus a Gaussian line.
The bottom panel shows the data to model ratio.
\label{figure:sp1}}
\end{figure}

The spectral fit is greatly improved ($\chi^2=144$ for 135 d.o.f.) by
using an absorbed power-law, which we associate with the nucleus. The
column density of cold gas toward the nucleus, inferred from 
fitting this model to the data, is significantly in excess of the
Galactic value, N$_H=4.4^{+1.8}_{-0.8} \times 10^{22}$ \nh, while
$\Gamma=1.40^{+0.09}_{-0.18}$. The measured flux of the absorbed power
law component is F$_{2-10~keV} \sim 6.4 \times 10^{-13}$ \flux, and
the intrinsic (absorption-corrected) luminosity is L$_{2-10~keV} \sim
8.5 \times 10^{40}$ \lum. Figure~\ref{figure:sp1} shows the EPIC data
with the model superposed. Line-like residuals are present in 0.7--0.9
keV, which could be due to an imperfect modelling of the thermal
emission from the halo (Gliozzi et al. 2003). 

Prompted by the results of the timing analysis, we investigated the
X-ray continuum in the ultra-soft energy range 0.3--0.8 keV. Adding
the lower-energy channels down to 0.3 keV requires an additional power
law component, with a fit improvement of $\Delta\chi^2$=11.5 for 1
additional dof. The photon index of the second power law is not well
determined, and assuming that ultra-soft X-ray emission comes from the
same region as the hard X-rays, it was linked to the index of the
primary power law. This ultra-soft component accounts for a
substantial fraction of the total counts in 0.3--0.8 keV (500 counts),
larger than systematic calibration uncertainties. A similar ultra-soft
non-thermal component is also required to fit the ACIS-S data (Gliozzi
et al. 2003). 

The \xmm\ observation confirms the previous result from \chandra\ that
excess X-ray absorption is present in NGC~4261 (Chiaberge et
al. 2002). These authors argue that the cold absorbing gas is located
in the nuclear dust lane visible in the \hst\ images (Jaffe et
al. 1996), if the dust-to-gas ratio in NGC~4261 is different from
Galactic. In fact, assuming a Galactic ratio the reddening implied by
the X-ray column is A$_{V, X} \sim 30$ mag, much larger than the
extinction in the dust lane, A$_{V,lane}=1.2-2.5$ mag (Chiaberge et
al. 2002). It is not unusual for Seyferts and other AGN to have X-ray
column densities in excess of those expected from the optical
reddening (e.g., Maiolino et al. 2001). 

To investigate the possibility of an ionized X-ray absorber, we fitted
the EPIC spectrum using a warm absorber ({\tt absori}), acting on the
power law only.  The column density N$_{W}$, ionization parameter
$\xi$, and temperature $T_W$, were left free to vary while the Fe
abundance was fixed to solar. We find an acceptable fit
($\chi^2=143/133$), although not better than the fit with the cold
absorber. The best-fit parameters, N$_{W}=5.5^{+0.5}_{-3.8} \times
10^{22}$ \nh, $\Gamma=1.30^{+0.22}_{-0.55}$, $\xi \sim 2.7$ erg cm
s$^{-1}$, and $T_W \sim 10^6$K, effectively mimic the case of a power
law plus a cold absorber. This is not unexpected, as the crucial
region 0.6--1 keV where most of the edges of ionized oxygen and other
elements would be present is masked by the thermal component,
preventing a meaningful determination of $\xi$. We thus conclude that
it is not possible to distinguish between a cold and warm absorber
with the EPIC data based on absorption properties.  
%However, one would expect a photoionized gas to emit soft X-ray
%lines. Interestingly, the EPIC data show line-like residuals in the
%range 0.7--0.9 keV (Fig. 3), suggesting Fe L emission which cannot be
%accounted for by emission from the thermal halo (Gliozzi et
%al. 2003). High-resolution X-ray observations could in principle
%discriminate photoionized emission lines from the lines due to thermal
%emission from a collisionally-ionized plasma. 

The connection between the X-ray and optical absorbing media in AGN
has been challenged by several authors (e.g., Weingartner \& Murray
2002). It is possible that the X-ray and optical absorbers are
separate gas components, with the X-ray absorbing gas lying inside the
large-scale obscuring torus, perhaps in the form of an ionized
wind. Such a scenario may be applicable to NGC~4261 as well as other
radio-loud AGNs, including some with broad optical emission lines
(e.g., 3C~445 and Arp~102B; see Sambruna et al. 1999).  Indeed, there
is evidence that in Arp~102B the absorbing X-ray and UV medium has the
form of an outflowing wind (Eracleous, Halpern, \& Charlton 2003). We
do note, however, that in broad-line objects the column density of the
X-ray absorber is an order of magnitude smaller than what is measured
in NGC~4261, which may indicate that in NGC~4261 we are observing the
combined effect of more than one absorbing component.

%Independent evidence for heavy absorption toward the nucleus of
%NGC~4261 (separate from the dust lane) is provided by photoionization
%models for its optical emission lines (Lewis, Eracleous, \& Sambruna
%2003). The line-emitting gas must receive a more intense UV flux than
%is directly observable (Zirbel \& Baum 1998) in order for the models
%to reproduce the observed emission-line ratios. This implies that our
%line of sight to the nucleus passes through 2.5--4.2 magnitudes of
%visual extinction, which is somewhat larger than the extinction that
%can be ascribed to the foreground dust lane.

\subsection{Detection of an Fe K emission line} 

Inspection of the residuals of the best-fit continuum component shows
a sharp line-like feature around 7 keV. Adding a Gaussian line to the
thermal+absorbed power law continuum model improves the fit to the
EPIC pn spectrum significantly, $\Delta\chi^2=7.52$ for 3 additional
parameters ($F$-test probability P$_F$ \gtsima 90\%). From the
confidence contours of the line normalization versus its width, we
derive that the line is detected (i.e., normalization larger than
zero) at \ltsima 95\% confidence, and always unresolved. A broad
($\sigma \sim$ 0.5 keV) line is not excluded at $\sim$ 90\%
confidence, with EW \ltsima 490 eV.

We thus fixed the rest-energy dispersion to the spectral resolution of
the EPIC pn in the Fe K range, $\sigma=130$ eV. (Fixing the line width
to much lower values does not change the results.)  To determine
if the Fe line is truly from the nuclear region, and not due to
an imperfect modelling of the spectrum of the diffuse halo, we
extracted EPIC pn spectra from increasingly larger radii and fitted
them with the same spectral model (Table 1). The Fe K line at 7 keV is
detected, and its significance increases with decreasing extraction radii. 

We conclude that the Fe line originates in the nucleus. Its best-fit
energy is consistent with the K$\alpha$ emission of H-like Fe (6.95
and 6.97 keV). Its measured EW of 280 eV is consistent with the value
measured by \asca\ (550$^{+300}_{-310}$ eV, Terashima et al. 2002),
and with the \chandra\ upper limit of 380 eV. An Fe K line at 7 keV is
also present in an archival 40 ks \sax\ observation, with EW $\sim
260$ eV (Gliozzi et al. 2003).

The measured EW of the line, although subject to large uncertainties,
is larger than the values measured for the \feka\ line in broad-line
radio-loud AGN at 6.4 keV, where EW $\sim$ 100 eV (e.g., Eracleous,
Sambruna, \& Mushotzky 2000; Wo\'zniak et al. 1998). In particular, it
is stronger than in sources where dilution from a beamed jet component
is thought to be at least partially responsible for the low EW of the
\feka\ line, e.g., 3C~120. Note, however, that even in broad lined objects
(as 3C~120 and 3C~390.3) there is a significant contribution to the X-ray 
continuum from the disk-corona based on spectral and timing properties
(Gliozzi, Sambruna \& Eracleous 2003).
As discussed above, 
either radiation from the inner jet is {\it not} the dominant origin
of the continuum emission in the nucleus of NGC~4261, or the X-ray
continuum is produced in the slower walls of the jet.

As the uncertainties on the line profile in the modest EPIC exposure
are large, we can only speculate on the line's origin. An origin in an
ionized disk (e.g., Ballantyne et al. 2002) is unlikely, as the
accretion rate in NGC~4261 is relatively low compared to the Eddington
rate.  The disk ionization parameter has a strong dependence on the
mass accretion rate, $\xi\propto \dot m^3$ where $\dot m=L/L_{\rm
Edd}$ (Matt, Fabian, \& Ross 1993).  To create H-like iron ions, $\dot
m > 0.5$ is required, while in NGC~4261 $L/L_{\rm Edd}\simeq 1\times
10^{-4}$ (Sambruna et al. 1999). 

A possible scenario for the formation of a narrow line is thermal
emission from an ADAF outflow (Narayan \& Raymond 2000). While
short-term X-ray flux variability rules out an ADAF as the {\it only}
mechanism of production of X-rays, it is still possible that an ADAF
is present in NGC~4261, contributing a fraction of the X-rays
including the Fe K line.
%Indeed, the measured photon index, $\Gamma=1.4$, is consistent
%with the value predicted for an ADAF (e.g., 
In this model, the Fe line originates from large radii and is thus
narrow, with expected EWs $\sim$ 200 eV. Also expected in this model
are emission lines of lighter elements at softer X-rays, including
highly ionized Ca and Ar in the energy range 3--4 keV, with EW of
several tens of eV, to which we are not sensitive. The ratios of the
line intensities would be a function of the outflow's mass and radius
(Narayan \& Raymond 2000).

However, the large uncertainties on the line width do not allow us to
exclude a broad line, or the presence of broad wings related to a
standard accretion disk. A deeper X-ray exposure is needed for more
meaningful constraints on the line profile. 

\subsection{NGC~4261 in perspective} 

It is interesting to compare NGC~4261 to more luminous radio-loud AGN
previously studied in X-rays (e.g., Sambruna et al. 1999). In Figure
3c of that paper we presented a plot of the 2--10 keV luminosity of
broad- and narrow-line radio-loud AGN versus their total 5 GHz
power. Extrapolating the correlation for broad-lined galaxies to the
lower X-ray luminosities, NGC~4261 is  
%Compared to the broad-lined radio galaxies, NGC~4261 is
overluminous by 1.5 decades in extended radio power. This raises the
interesting possibility that the faint nuclear emission from this
galaxy is due to a very efficient channeling mechanism of its
accretion power into the jets and the radio lobes. A similar scenario
was indeed proposed for the FRI galaxy IC4296 (Pellegrini et
al. 2002), which also hosts a LINER nucleus.  In IC4296, the accretion
rate would not be different from the Bondi accretion rate and the jet
would account for a large fraction of the accretion power. The LINER
spectra of both sources would thus be due to a diminished, isotropic
optical-to-UV luminosity. 

Whether or not an ADAF or a very efficient channeling mechanism (or a
combination of the two) is present in NGC~4261, it is clear that the
transition from high to low nuclear luminosities in radio-loud AGN
corresponds to a change of nuclear properties. Because of its well
known characteristics (black hole mass, inclination, pc-scale jet),
NGC~4261 has a key role to play in the study this transition.
                        
\acknowledgements 

Financial support from NASA LTSA grants NAG5-10708 (MG, RMS),
NAG5-8107 (WNB), and NAG5-9982 (ME) is gratefully acknowledged. Funds
were also provided by NASA grant NAG5-10243 (MG, RMS) and by the Clare
Boothe Luce Program of the Henry Luce Foundation (RMS).

%%%%%%%%%%%%%%%%%%%%%%%%%%%%%%%%%%%%%%%%%%%%%%%%%%%%%%%%%%%%%%%%%%%%%%%%%%
%%%%%%%%% R E F E R E N C E S
%%%%%%%%%%%%%%%%%%%%%%%%%%%%%%%%%%%%%%%%%%%%%%%%%%%%%%%%%%%%%%%%%%%%%%%%%%

%\clearpage

%------------------------------------------------------------------------
% TABLE 1
%------------------------------------------------------------------------

\begin{deluxetable}{llll}
\tablenum{1}
\tablewidth{4.2in}
\tablecolumns{4}
\tablecaption{Fits to the EPIC pn spectrum of NGC~4261}
\tablehead{
\colhead{Parameter$^{~a}$} &
\colhead{r=10"} &
\colhead{r=20"} &
\colhead{r=30"} 
}
\startdata
kT (keV) & $0.65^{+0.01}_{-0.02}$ & $0.66^{+0.01}_{-0.01}$ &
$0.67^{+0.01}_{-0.01}$ \\ 
Abundance  & $0.36^{+0.12}_{-0.07}$ & $0.44^{+0.63}_{-0.12}$ &
$0.33$ \\ 
$N_H ~(10^{22} cm^{-2})$  & $4.4^{+1.8}_{-0.8}$ & $3.0^{+1.5}_{-1.8}$ &
$2.8^{+1.3}_{-0.9}$\\
Photon Index $\Gamma$  & $1.40^{+0.09}_{-0.18}$ & $1.30^{+0.30}_{-0.29}$ &
$1.25^{+0.07}_{-0.08}$  \\ 
$E_{\rm line}$ (keV) & $7.00^{+0.09}_{-0.08}$ & $6.96^{+0.11}_{-0.12}$ &
$6.98^{+0.20}_{-0.29}$   \\ 
EW (eV) & $281^{+200}_{-164}$ & $264^{+187}_{-172}$ &
$164^{+161}_{-156}$  \\ 
$\chi^2/d.o.f.$& 136.50/133 & 221.23/185  &
251.12/212\\ 
%$L_{2-10~keV}~(10^{40}{\rm erg~s^{-1}})$ &8.5 & 7.7 & 7.8\\
$P_{\rm Ftest}(\%)^{~b}$ & 97 & 92 & 76\\
\enddata
\tablenotetext{a\;}{Errors on the spectral parameters are 90\%
confidence for one parameter of interest ($\Delta\chi^2$=2.7).}
\tablenotetext{b\;}{Probability from the F-test that the 
addition of the Gaussian line to the continuum best-fit model is
significant.}  
\end{deluxetable}

%% \end{document}

%%*************** X-RAY EPIC MOS1 & CHANDRA ACIS S NGC~4261 ************

%%%*************** X-RAY EPIC pn lc NGC~4261 r=10"*******************************

%%%***************NGC~4261 SPECTRUM**************************************

%%%%%%%%%%%%%%%%%%%%%%%%%%%%%%%%%%%%%%%%%%%%%%%%%%%%%%%%%%%%%%%%%%%%%%%%%%
%%%%%%%%% T H E   E N D 
%%%%%%%%%%%%%%%%%%%%%%%%%%%%%%%%%%%%%%%%%%%%%%%%%%%%%%%%%%%%%%%%%%%%%%%%%%

\end{document}